\documentclass[twocolumn,twocolappendix]{aastex631}
\usepackage[T1]{fontenc}
\usepackage[utf8]{inputenc}
\usepackage[english]{babel}
\usepackage{amssymb}
\usepackage{amsmath}
\usepackage{esint}
\usepackage{bm}
\usepackage{hyperref}
\usepackage{bookmark}
\usepackage{xcolor}
\usepackage{textgreek}

\DeclareMathAlphabet{\mathup}{OT1}{\familydefault}{m}{n}
\newcommand\diff{\mathop{}\!\mathup{d}}

\newcommand\tsup[1]{\textsuperscript{#1}}

\newcommand\supn[1]{\tsup{\ensuremath{#1}}}

\makeatletter
\newcommand\DeclareUnit[2]{%
    \@namedef{#1}{\@ifnextchar[{\csname @with@#1\endcsname}{\csname @without@#1\endcsname}}%
    \@namedef{@with@#1}[##1]{\text{#2\supn{##1}}}%
    \@namedef{@without@#1}{\text{#2}}%
}%
\makeatother

\DeclareUnit{g}{\text{g}}
\DeclareUnit{cm}{\text{cm}}
\DeclareUnit{mm}{\text{mm}}
\DeclareUnit{um}{\text{\textmu m}}
\DeclareUnit{au}{\text{au}}

\newcommand{\planet}{\text{pl}}
\newcommand{\MEarth}{\ensuremath{\mathup{M}_\oplus}}
\newcommand{\MJup}{\ensuremath{\mathup{M}_\text{Jup}}}
\newcommand{\MSun}{\ensuremath{\mathup{M}_\odot}}

\newcommand{\gas}{\text{gas}}
\newcommand{\dust}{\text{dust}}
\newcommand{\grain}{\text{gr}}
\newcommand{\Reynolds}{\mathop{}\!\mathup{Re}}
\newcommand{\Stokes}{\mathop{}\!\mathup{St}}
\newcommand{\dragcoeff}{\ensuremath{C_d}}

\begin{document}
\title{Dynamics of small, constant size particles in a protoplanetary disk with an embedded protoplanet}

\author[0000-0002-3286-3543]{Ellen M. Price}
\altaffiliation{Heising-Simons Foundation 51 Pegasi b Postdoctoral Fellow}
\affiliation{Department of the Geophysical Sciences, University of Chicago, 5734 South Ellis Ave., Chicago, IL 60637, USA}

\author[0000-0002-5954-6302]{Eric Van Clepper}
\affiliation{Department of the Geophysical Sciences, University of Chicago, 5734 South Ellis Ave., Chicago, IL 60637, USA}

\author[0000-0002-0093-065X]{Fred J. Ciesla}
\affiliation{Department of the Geophysical Sciences, University of Chicago, 5734 South Ellis Ave., Chicago, IL 60637, USA}

\begin{abstract}
Hydrodynamical simulations of protoplanetary disk dynamics are useful tools for understanding the formation of planetary systems, including our own. Approximations are necessary to make these simulations computationally tractable. A common assumption when simulating dust fluids is that of a constant Stokes number, a dimensionless number that characterizes the interaction between a particle and the surrounding gas. Constant Stokes number is not a good approximation in regions of the disk where the gas density changes significantly, such as near a planet-induced gap. In this paper, we relax the assumption of constant Stokes number in the popular FARGO3D code using semi-analytic equations for the drag force on dust particles, which enables an assumption of constant particle size instead. We explore the effect this change has on disk morphology and particle fluxes across the gap for both outward- and inward-drifting particles. The assumption of constant particle size, rather than constant Stokes number, is shown to make a significant difference in some cases, emphasizing the importance of the more accurate treatment.
\end{abstract}


\section{Introduction}

\noindent Chondritic meteorites are generally believed to contain the most pristine remaining material from the protosolar disk, and their study may well reveal details about the formation of the Solar System. Meteorite classification has changed over time, but \citet{Warren2011} suggests a simple distinction of meteorites, based on the abundances of several stable isotopes, into two classes: carbonaceous chondrites (CC) and non-carbonaceous chondrites (NC). \citet{Warren2011} notes a ``striking bimodality'' in distributions of stable isotope ratios among these classes with no known samples falling in between them. Similar isotopic variations suggesting at least two distinct material reservoirs have been found since \citep[e.g.,][]{Moynier+2012,FuriMarty2015,Yokoyama+2015,Budde+2016,Bermingham+2018}. Since the first identification of the distinct CC and NC groups, the origin of the bifurcation in meteorite composition has been an open question.

If accretion in two distinct material reservoirs explains these two classes of meteorites, then one promising theory is that the forming Jupiter's orbit served as a hard barrier to mixing between the reservoirs \citep[e.g.,][]{Kruijer+2017,Budde+2018,Nanne+2019,Kruijer+2020}, blocking inward radial drift of solids from the outer disk in particular. However, some models suggest that this simple picture is incomplete. \citet{Liu+2022} raise an inconsistency with the theory, arguing that the inner NC reservoir is too rapidly depleted by radial drift without some material inflow from the outer disk; they propose an alternative model that invokes a combination of drift and viscous spreading to explain the compositions of meteorites. Other recent numerical simulations show that the Jupiter gap may not be an efficient barrier when fragmentation of solids occurs, grinding large pebbles down to dust that can be transported through the gap by drift or diffusion \citep{Drazkowska+2019,Stammler+2023}. Outward drift is not addressed as frequently in the literature, but two recent studies \citep{Schrader+2020,Schrader+2022} find that some CC chondrule compositions are consistent with NC material, suggesting drift across the Jupiter gap to the outer disk and subsequent mixing may have occurred in the early Solar System.

Many hydrodynamical simulations of how a Jupiter-carved gap impacts small, drifting solids have been carried out to date, with varying degrees of complexity. \citet{Eriksson+2020} solve the nonlinear diffusion equation for a disk's surface density as it is perturbed by a protoplanet, a one-dimensional approach suggested by \citet{LinPaploizou1986}. While computationally convenient, one-dimensional, azimuthally-averaged evolution equations over-simplify the full nature of the interaction between a protoplanet and its disk. In earlier work, \citet{Paardekooper+2004} simulate, in two dimensions, the opening of a gap by a 0.1~\MJup\ planet in a disk of dust and gas fluids, using 1~mm particles assumed to be in the Epstein drag regime; a similar study is carried out by \citet{Paardekooper+2006} to investigate dust and gas dynamics in the presence of an embedded planet. \citet{Paardekooper2007} relax the assumption of a continuous dust fluid and carry out two-dimensional simulations of a gas fluid and discrete dust particles, influenced by both Epstein and Stokes drag laws, to measure accretion onto a massive embedded planet.

Moving to three-dimensions, \citet{Maddison+2007} and \citet{Fouchet+2007} develop a smoothed-particle hydrodynamics simulation of a dusty gas fluid, assuming particles stay in the Epstein drag regime; \citet{Fouchet+2007} find, unlike previous two-dimensional studies, that the width and depth of the planet-induced gap depends on grain size, a difference they attribute to varying scale heights with particle size. The work of \citet{Ayliffe+2012} agrees with previous findings \citep[e.g.,][]{Fouchet+2007,Paardekooper2007,Lyra+2009}, suggesting that planet formation may occur rapidly in regions where particles accumulate outside planet-induced gaps. \citet{Binkert+2021} moves to using a three-dimensional, grid-based approach, returning to the dust fluid (continuum rather than particle-based) prescription of earlier studies and assuming Epstein drag. They find that the time-dependent dust density structure differs significantly from the gas density structure, pointing to larger disk mass estimates than previously suggested from ALMA results. All of these studies reinforce and refine our understanding of the complex interplay of dust and gas dynamics and disk substructure.

While two- and three-dimensional models do have the necessary complexity to capture dynamics between gas and solid particles in the presence of an embedded planet, they may miss important features if they fix the particle Stokes number rather than particle size. Some hydrodynamics codes already implement fixed size particle drag forces, including Phantom \citep{Price+2018}, PLUTO \citep{Mignone+2019}, Athena++ \citep{Huang+2022}, and RAMSES \citep{Moseley+2023}. Others, however, use fixed Stokes number particles, which are computationally convenient but not as physically-motivated as particles of fixed size. For example, \citet{Pierens+2019} do this, as does the popular FARGO3D code \citep{fargo3d,fargo3dmultifluid}, as used by \citet{Sturm+2020} and \citet{Chan+2024}. Given the rich history of increasing nuance with increasing complexity in models of protoplanetary disks with embedded planets, this treatment may or may not be justified; we aim to investigate the differences that arise between simulations of fixed size and fixed Stokes number particles, to aid in the interpretation of work done to date.

The Stokes number is a dimensionless number commonly encountered in fluid dynamics when studying the interaction between a fluid and solid particles suspended in it. In protoplanetary disks, it follows the relation
\begin{equation}
    \Stokes \propto \frac{a_\grain \rho_\grain}{\Sigma_\gas},
    \label{eqn:stokesdef}
\end{equation}
where $a_\grain$ and $\rho_\grain$ are the radius and density, respectively, of a single dust grain, and $\Sigma_\gas$ is the gas surface density. Since the Stokes number depends on particle size and local gas properties, the dynamics around large shifts in gas density may change depending on the assumptions made about the solids; holding the Stokes number fixed while decreasing the surface density is equivalent to spontaneously increasing the grain size.

Several attempts to relax the approximation of constant Stokes number have been made in other works; we summarize those here to distinguish our treatment from theirs. \citet{Weber+2018} report surface density evolution for particles of constant size using scaling relations. \citet{Auddy+2022} derive a set of steady-state drift equations for particles of constant size (their Appendix~A), which they implement in FARGO3D. \citet{Dullemond+2022} take a different approach and numerically integrate fixed size particle trajectories using precomputed gas dynamics from PLUTO \citep{pluto}, which they compare to similar results from FARGO3D. In addition to being limited to a finite number of particles, this approach excludes the possibility of feedback from the dust onto the gas, since the solids are only introduced in post-processing. \citet{Wu+2023} modify FARGO3D's source code to allow fixed size particles, though they do not provide detail about how this was accomplished. Here, we utilize analytic equations for the evolution of the solid particles' momenta, derived in Appendix~\ref{adx:analytic}, which are exact within a timestep, allow feedback between solids and gas, and can be readily integrated into any hydrodynamics code\footnote{We make our implementation of fixed size particle drag in FARGO3D available at \url{https://github.com/emprice/fargo3d/tree/feature/fixed-size-drag}.}.

The goal of this paper is to explore the dynamics of dust grains in the presence of a massive, embedded protoplanet and investigate the differences that arise when treated as particles of constant size, rather than constant Stokes number. In the context of the meteorite dichotomy, we seek to characterize any mixing that occurs between the material reservoirs inside and outside the gap formed by the protoplanet, considering both inward drift from the outer disk and outward drift from the inner disk.

In Section~\ref{sec:methods}, we present our modifications to the methods of \citet{fargo3dmultifluid} that allow us to make this comparison. In Section~\ref{sec:results}, we present the results of our analysis. We discuss and conclude in Sections~\ref{sec:discussion} and \ref{sec:conclusions}, respectively. Supporting mathematical derivations are provided in Appendix~\ref{adx:analytic}, and supplemental figures can be found in Appendix~\ref{adx:suppfig}.

\section{Methodology}
\label{sec:methods}

\noindent In protoplanetary disks, gas molecules experience an additional momentum flux due to gas pressure that solids do not. In the presence of a central star and no other external forces, the gas of a protoplanetary disk, in perfect equilibrium, orbits at a sub-Keplerian velocity, so long as the pressure gradient satisfies $\diff p / \diff r < 0$. Solid particles, on the other hand, would orbit at exactly Keplerian velocity, and so, in the frame of a solid particle, the bulk gas surrounding it exerts a headwind. Assuming elastic collisions between the particle and gas molecules, the total momentum is conserved, but momentum is exchanged between the two phases. The behavior of a mixture of gas and dust, then, is not trivial to model once the drag force on the dust is included, since the dynamics of both species are influenced by the momentum exchange. In a mixture that includes multiple dust species, all of them may exchange momentum with the gas and, indirectly, with each other, further coupling the physics of all the mixture's components.

\subsection{Simulation setup}
\label{sec:setup}

\noindent We employ the FARGO3D code \citep{fargo3d,fargo3dmultifluid} to simulate a protoplanetary disk with an embedded protoplanet as a multifluid mixture of a bulk gas and five dust species in two dimensions. In the code and throughout this paper, we use cylindrical coordinates $\left(r, \varphi\right)$, where $r$ is the radial coordinate and $\varphi$ is the azimuthal angle. The common model parameters are given in Table~\ref{tbl:parameters}.

For the initial conditions, we adopt the following. The initial surface densities of the gas and dusts are given by
\begin{equation}
    \Sigma_\gas\!\left(t = 0, r, \varphi\right) = \Sigma_0 \left(\frac{r}{r_0}\right)^{-\beta}
\end{equation}
and
\begin{equation}
    \Sigma_\dust\!\left(t = 0, r, \varphi\right) = \epsilon \Sigma_\gas\!\left(t = 0, r\right),
\end{equation}
respectively. The radial velocities of all fluids are initially zero; the initial azimuthal velocities of the gas and dust are
\begin{equation}
    v_{\varphi,\gas}\!\left(t = 0, r, \varphi\right) = \Omega r \sqrt{1 - \psi^2 \left(\beta + 1\right)}
\end{equation}
and
\begin{equation}
    v_{\varphi,\dust}\!\left(t = 0, r, \varphi\right) = \Omega r.
\end{equation}
The speed of sound is fixed in time in our simulations but varies radially as
\begin{equation}
    c_s\!\left(r\right) = \Omega r \psi.
\end{equation}
We adopt the typical definition of the orbital frequency, $\Omega^2 \equiv G M_\star / r^3$, and $\psi$ is the constant disk aspect ratio. Since our model is two-dimensional, but the drag equations in Appendix~\ref{adx:analytic} depend on the volume density of gas and dust, we must make some assumption about the vertical structure of the disk. We choose the disk to have a vertical Gaussian density distribution with an integrated value equal to the surface density, so
\begin{equation}
    \rho\!\left(t, r, \varphi, z = 0\right) = \frac{\Sigma\!\left(t, r\right)}{\sqrt{2\pi} h}
\end{equation}
is the volume density at the disk midplane, with scale height $h \equiv c_s / \Omega$.

Rather than introducing a planet into the disk instantaneously, which might lead to shocks or extreme oscillations, we use FARGO3D's built-in mass taper function to increase the planet mass gradually with time according to
\begin{equation}
    M_\planet\!\left(t\right) = M_\planet \times \begin{cases}
        \frac{1}{2} \left[1 - \cos{\frac{\pi t}{\tau_\planet}}\right] & t < \tau_\planet \\
        1 & t \geq \tau_\planet
    \end{cases},
\end{equation}
where $M_\planet$ is the final planet mass and $M_\planet\!\left(t\right)$ is the value used in computing the gravitational potential.

Our fiducial simulation resolution is given in Table~\ref{tbl:parameters}. Because low values of $\alpha$ can trigger the Rossby wave instability (RWI; e.g., \citealt{Chan+2024}), we have chosen to use a relatively high value of $10^{-3}$, which ensures that simulations run at higher resolution give the same qualitative results.

\begin{deluxetable}{lcr}
    \tablecaption{Model parameters}
    \tablehead{\colhead{Name} & \colhead{Symbol} & \colhead{Value}}
    \startdata
        Inner disk radius & $r_\text{min}$ & 0.4$r_0$ \\
        Outer disk radius & $r_\text{max}$ & 2.5$r_0$ \\
        Mesh resolution in $r$ & $N_r$ & $128$ \\
        Mesh resolution in $\varphi$ & $N_\varphi$ & $384$ \\
        \hline
        Normalization radius & $r_0$ & 5.2~\au \\
        Normalization surface density & $\Sigma_0$ & $32.9$~\g~\cm[-2] \\
        Initial surface density slope & $\beta$ & $1/2$ \\
        Disk aspect ratio & $\psi$ & 0.05 \\
        Stellar mass & $M_\star$ & $1$~\MSun \\
        Viscosity parameter\tablenotemark{\scriptsize a} & $\alpha$ & $10^{-3}$ \\
        Dust-to-gas ratio & $\epsilon$ & 0.01 \\
        \hline
        Planet mass growth time & $\tau_\planet$ & 500~orbits \\
        Planet orbital radius & $r_\planet$ & $1 r_0$ \\
        \hline
        Solid density & $\rho_\grain$ & 3~\g~\cm[-3] \\
        Particle size\tablenotemark{\scriptsize b} & $a_1$ & 0.1~\um \\
        & $a_2$ & 1~\um \\
        & $a_3$ & 10~\um \\
        & $a_4$ & 100~\um \\
        & $a_5$ & 1~\mm \\
        \hline
        Particle Stokes number\tablenotemark{\scriptsize c} & $\mathrm{St}_1$ & $2 \times 10^{-6}$ \\
        & $\mathrm{St}_2$ & $2 \times 10^{-5}$ \\
        & $\mathrm{St}_3$ & $2 \times 10^{-4}$ \\
        & $\mathrm{St}_4$ & $2 \times 10^{-3}$ \\
        & $\mathrm{St}_5$ & $2 \times 10^{-2}$ \\
    \enddata
    \tablenotetext{\scriptstyle a}{Assuming the \protect\citet{ShakuraSunyaev1973} $\alpha$ viscosity model where kinematic viscosity $\nu = \alpha c_s^2 / \Omega$}
    \tablenotetext{\scriptstyle b}{When particle size is fixed}
    \tablenotetext{\scriptstyle c}{When particle Stokes number is fixed}
    \label{tbl:parameters}
\end{deluxetable}

\subsection{Dust drag for particles of fixed size}
\label{sec:fixedsize}

\noindent By default, FARGO3D computes the drag force on dust grains by assuming a fixed Stokes number. The numerical scheme used by FARGO3D is stable if the Stokes number varies spatially \citep{fargo3dmultifluid}, but the public version of the code does not currently include a mechanism for prescribing a local Stokes number. The Stokes number is proportional to the stopping time of the dust \citep[e.g.,][]{Birnstiel+2010}, which scales inversely with $\rho_\gas$, a quantity that changes by orders of magnitude across a gap cleared by a planet. To investigate how holding the dust size fixed may change the dynamics and evolution of a simulated disk, we develop an extension to the FARGO3D code that computes the drag force consistent with a fixed particle size.

In general, solving the full set of coupled differential equations that govern the drag forces across the Stokes and Epstein regimes requires solving an initial value problem at every timestep and for every computational cell. We find that using an integrator for this purpose is prohibitively slow and ultimately unnecessary. FARGO3D takes small timesteps by design, so initializing and running robust integration software over just a small $\Delta t$ at every timestep can increase the simulation wall time dramatically. As an alternative, if we assume that feedback between dust and gas is negligible except for the largest of the dust species, the drag equations have analytic solutions, listed in Appendix~\ref{adx:analytic}, which are exact and can be computed very efficiently.

Table~\ref{tbl:parameters} lists the Stokes numbers and particle sizes used in our simulations. From Equation~\ref{eqn:stokesdef}, it is impossible to choose exactly one Stokes number that will always correspond to exactly one particle size. The particle size bins and Stokes number bins were chosen to roughly correspond over most of the outer disk, and they are not expected to coincide for all radii or all times.

\subsection{Computing particle trajectories}

\noindent FARGO3D produces as output time-varying density and velocity fields for each fluid in the simulation. Obtaining the trajectory of a particle with a given initial position can be accomplished straightforwardly by solving the initial value problem
\begin{equation}
    \begin{pmatrix}
        \dot{r} \\
        \dot{\varphi}
    \end{pmatrix} = \begin{pmatrix}
        v_r\!\left(t, r, \varphi\right) \\
        v_\varphi\!\left(t, r, \varphi\right) / r
    \end{pmatrix},
\end{equation}
where the radial and azimuthal velocity fields ($v_r$ and $v_\varphi$, respectively) are smoothly interpolated in time and position from the FARGO3D output. We \textit{do not} add contributions from gravitational or drag forces during this integration, as they are already taken into account in the FARGO3D simulation.

We distribute 1000 test particles uniformly in an annulus in either the inner disk or outer disk, somewhat away from $r = r_0$, so that the particles' dynamics can transport them across the gap. Particles seeded in the inner disk are initialized with $0.65 < r < 0.85$, and those seeded in the outer disk are initialized with $1.15 < r < 1.35$; the angular distribution is uniform and offset by the planet's angular position so that different simulations can be compared fairly. To investigate the effect of the time particles are released into the disk, we perform these trajectory computations at 200~orbits, 400~orbits, 600~orbits, and 800~orbits, with a total simulation time of 2000~orbits (equivalent to 3780~years). Particle release times vary between different studies, and ours are of the same order of magnitude as those in \citet{Binkert+2023}, but they do not guarantee a steady state has been reached.

\section{Results}
\label{sec:results}

\noindent Below, we consider two different ways to compare the effects of fixing particle size rather than Stokes number. First, we examine the overall disk morphology, which depends on variations in the dust fluid evolution, at a fixed time. Then, we evolve step function surface density profiles from the inner disk or outer disk to measure how material from one reservoir crosses to the other.

\subsection{Disk morphology}
\label{sec:morphology}

\begin{figure*}
    \centering
    \includegraphics[width=\linewidth]{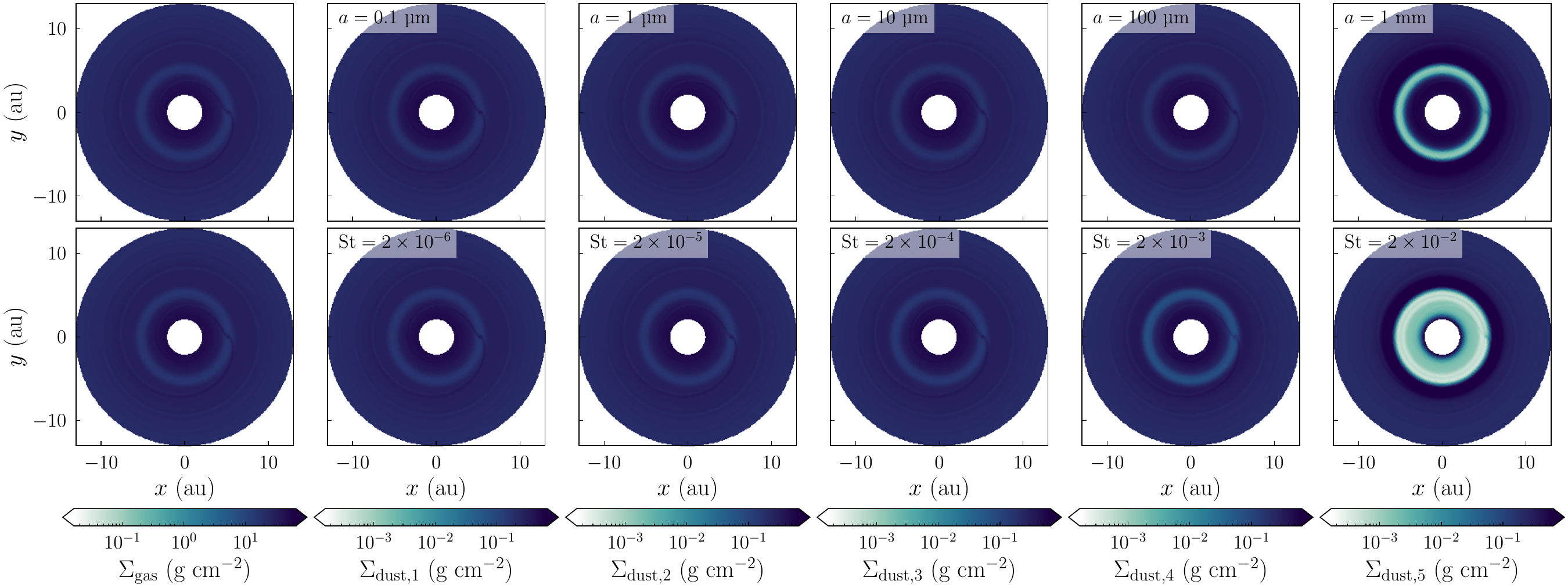}
    \caption{Surface densities of the simulated disk after 2000~orbits in each of the six fluids (the gas fluid and five dust fluids). The planet lies along the positive $x$-axis. Note that the surface density scale differs by a factor $10^2$ between the gas and the dust species, which reflects the initial dust-to-gas ratio. The final mass of the planet is $M_\planet = 30~\MEarth$.}
    \label{fig:morphology-30earth}
\end{figure*}

\begin{figure*}
    \centering
    \includegraphics[width=\linewidth]{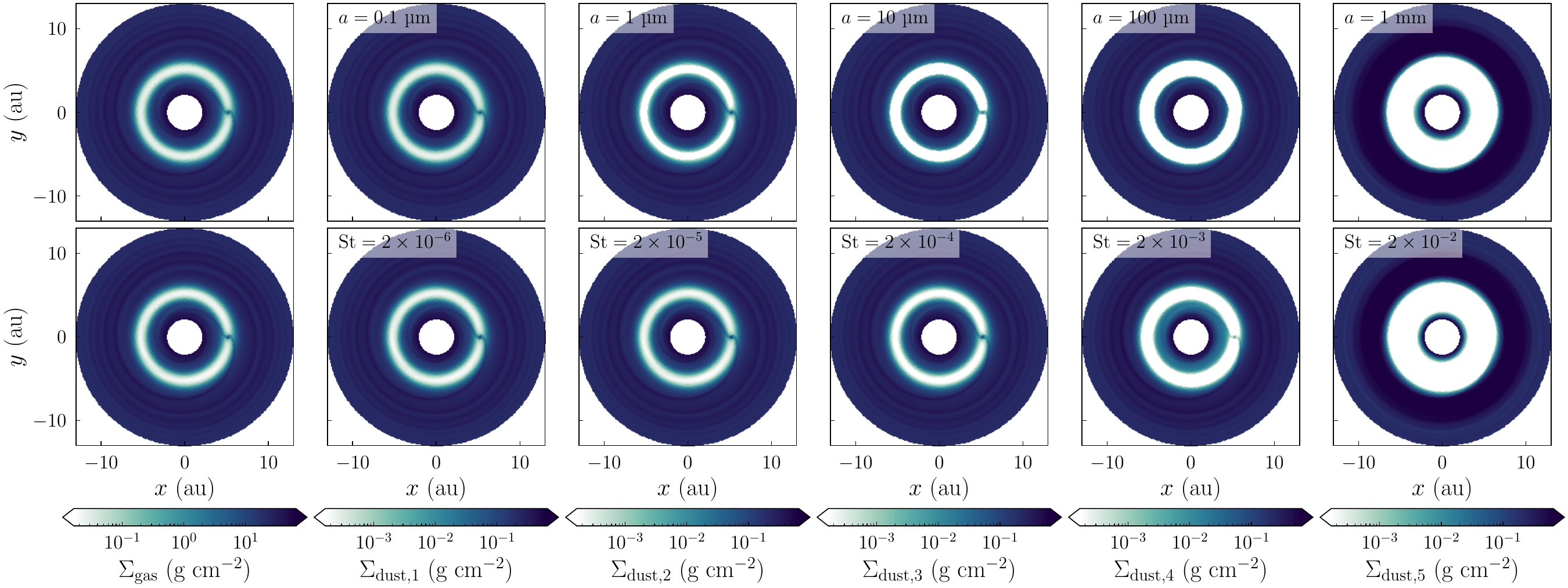}
    \caption{Same as Figure~\ref{fig:morphology-30earth}, but for a final planet mass $M_\planet = 1~\MJup$.}
    \label{fig:morphology-jupiter}
\end{figure*}

\begin{figure}
    \centering
    \includegraphics[width=\linewidth]{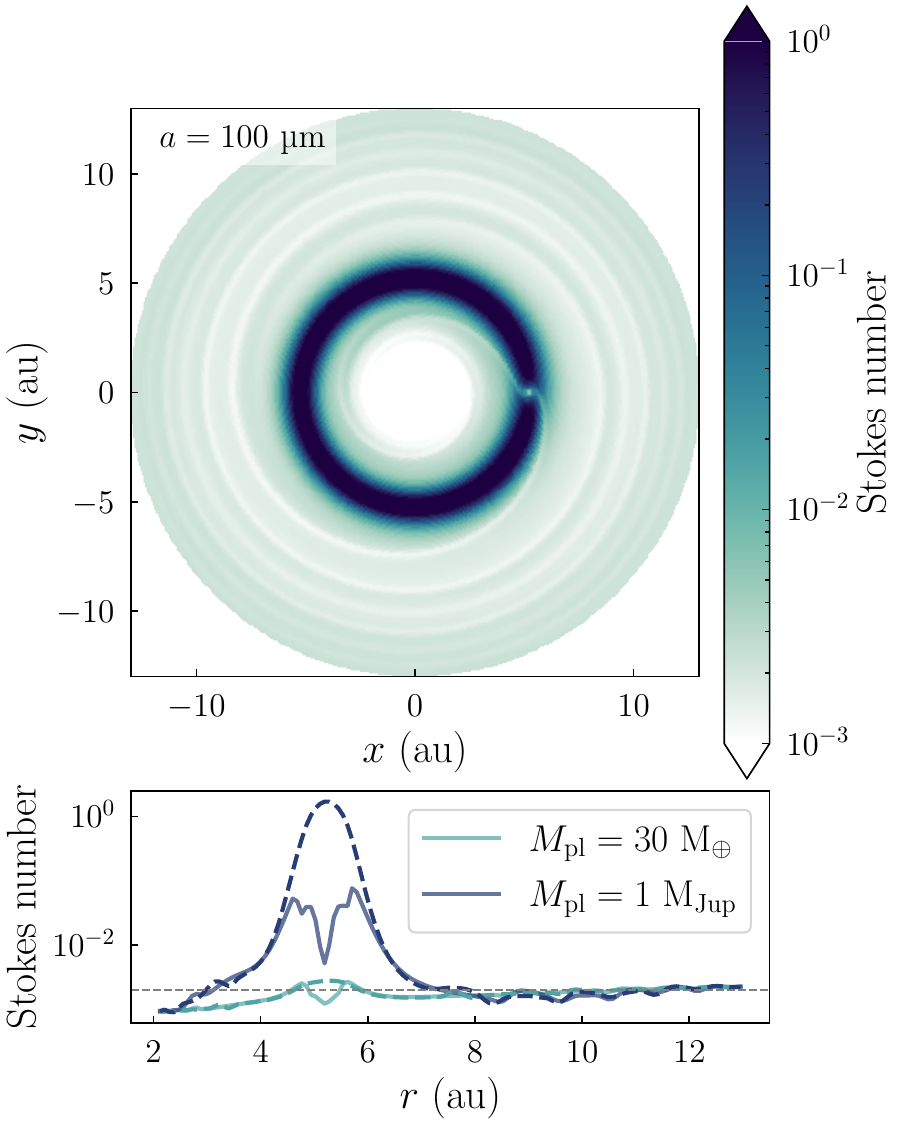}
    \caption{In the upper panel, we show a map of the Stokes number computed for particles of constant size 100~\um\ in the presence of  a Jupiter-mass planet, after evolution for 2000~orbits. In the lower panel, we show the Stokes number along negative $x$-axis (dashed) and positive $x$-axis (solid) for two different planet masses, indicating the corresponding assumed Stokes number with a dashed line. While the Stokes number increases dramatically inside the planet-induced gap compared to its surroundings, there is little to no material there to experience this level of decoupling from the gas. In the outer disk, we see that variations in the Stokes number are of order a few.}
    \label{fig:stokes-map-jupiter}
\end{figure}

\noindent In Figures~\ref{fig:morphology-30earth} and \ref{fig:morphology-jupiter}, we show dust and gas surface densities of a simulated disk after 2000~orbits, for a 30~\MEarth\ and 1~\MJup\ planet, respectively. These surface density profiles provide a quick way to compare the outcomes for fixed size and fixed Stokes number particles. Immediately, we observe that the disk morphologies in the 1~\MJup\ case are qualitatively similar, but there are more obvious differences in the 30~\MEarth\ case. The largest fixed Stokes number particles (with $\Stokes = 0.02$) are drained more efficiently from the inner disk, leaving a lower-density region interior to the planet's orbit, than the corresponding fixed size particles (with $a = 1~\mm$). In Figure~\ref{fig:stokes-map-jupiter}, we show a map of the Stokes number for a single particle size, to demonstrate the magnitude of the variations and their spatial dependence. Based on Figure~\ref{fig:stokes-map-jupiter}, it becomes clear that the fixed Stokes number approximation is better in the outer disk than in the inner disk, where the measured Stokes number of fixed size dust is actually lower when $M_\planet = 30~\MEarth$. This explains the efficient draining of the fixed $\Stokes = 0.02$ material in the inner disk in Figure~\ref{fig:morphology-30earth}: The fixed Stokes number dust there is more decoupled from the gas and susceptible to rapid inward drift.

We additionally provide Figures~\ref{fig:morphology-30earth-zoom} and \ref{fig:morphology-jupiter-zoom} that are zoomed in to show more detail around the gap in Figures~\ref{fig:morphology-30earth} and \ref{fig:morphology-jupiter}, respectively; the largest differences are observed around the planet-induced gap edges, where the density gradients are strongest.

\subsection{Step function evolution}
\label{sec:stepfn}

\begin{figure}
    \centering
    \includegraphics[width=\linewidth]{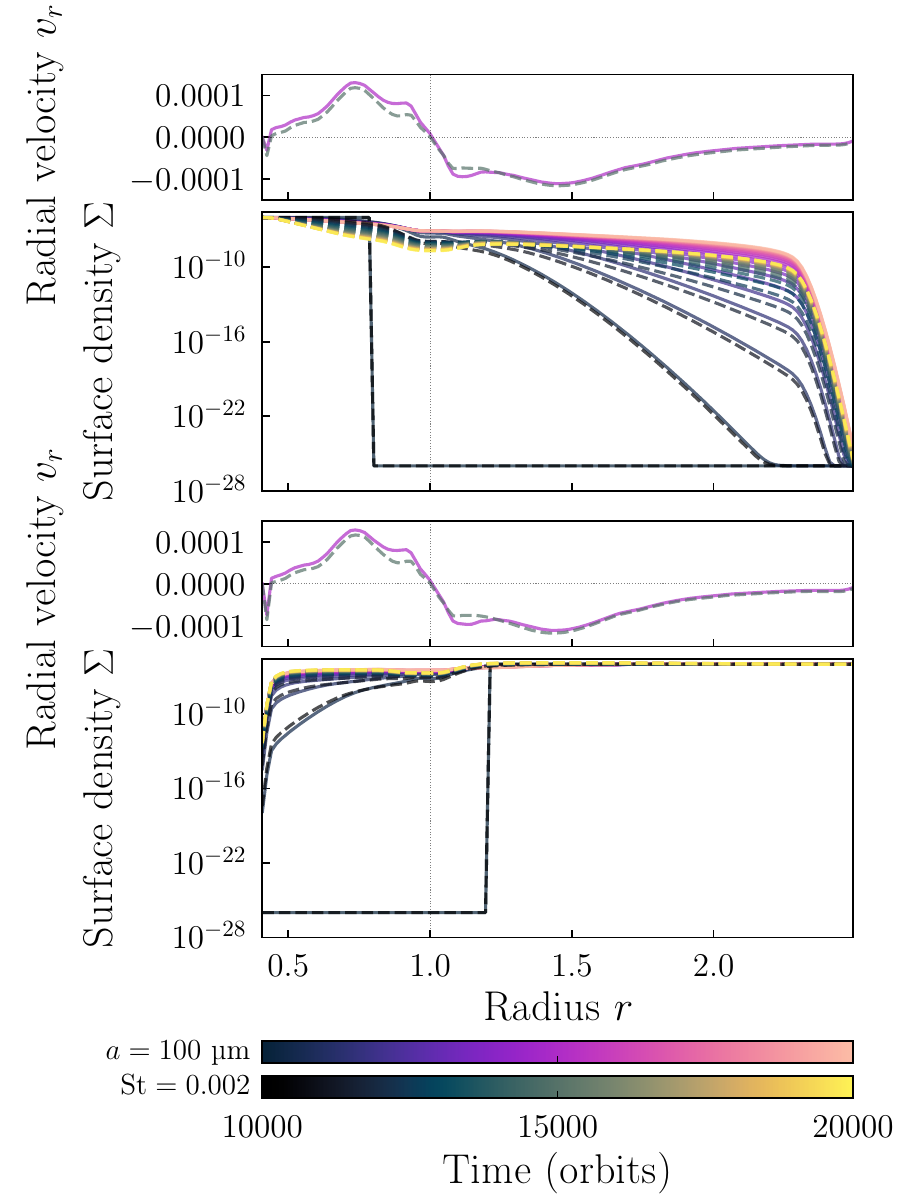}
    \caption{Evolution of the azimuthally-averaged dust radial velocity and surface density in a disk with a 30~\MEarth\ planet when the initial dust profile is a step function introduced after $10^4$ orbits. In the upper two panels, virtually all dust starts in the inner disk; the surface density outside $r = 0.8$ is negligible but nonzero for numerical stability. In the lower two panels, virtually all dust starts in the outer disk, outside $r = 1.2$. Solid lines (in pink and purple color) correspond to fixed size particles with $a = 100$~\um, and dashed lines (in blue and yellow color) correspond to fixed Stokes number particles of $\Stokes = 2 \times 10^{-3}$. All quantities are plotted in code units.}
    \label{fig:stepfn-30earth-100um}
\end{figure}

\begin{figure}
    \centering
    \includegraphics[width=\linewidth]{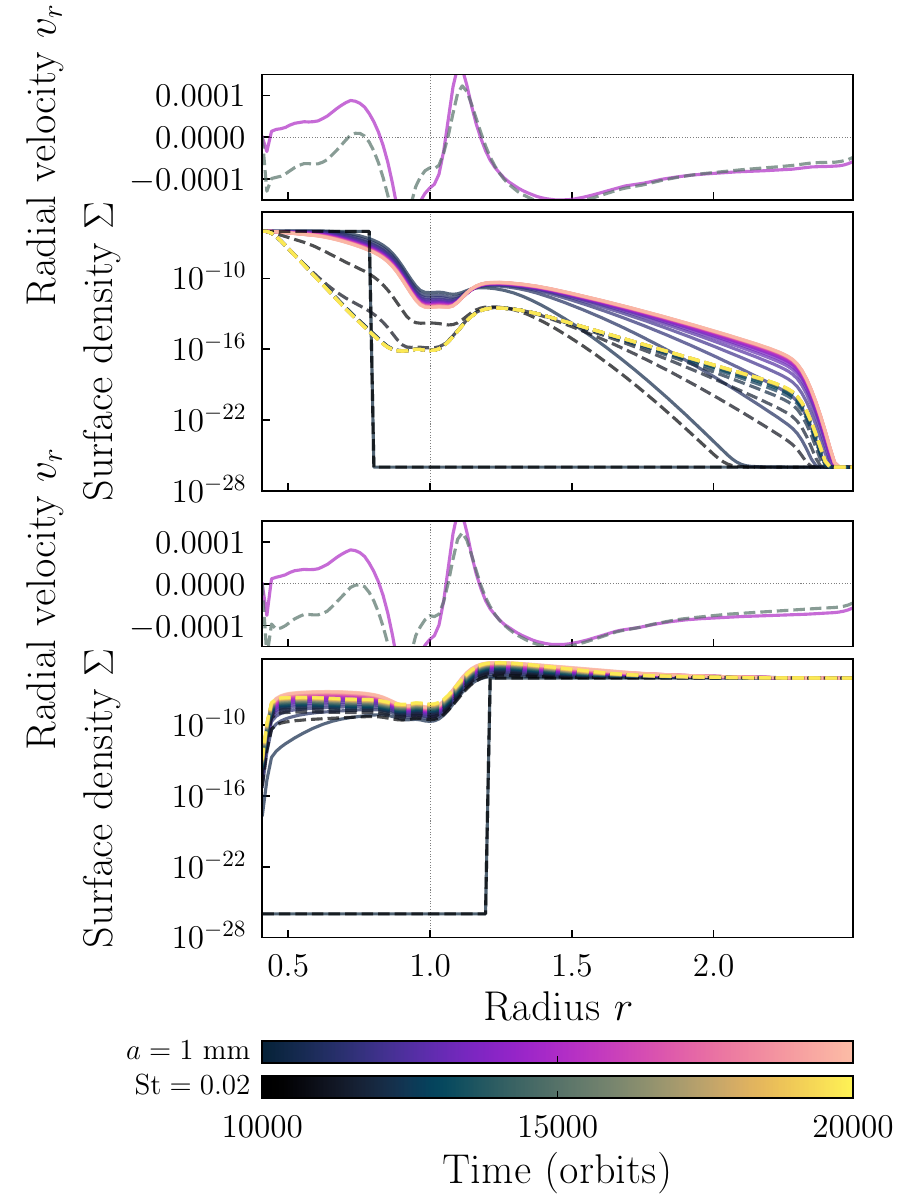}
    \caption{Same as Figure~\ref{fig:stepfn-30earth-100um}, but for $a = 1$~\mm\ (solid lines) and $\Stokes = 2 \times 10^{-2}$ (dashed lines).}
    \label{fig:stepfn-30earth-1mm}
\end{figure}

\begin{figure}
    \centering
    \includegraphics[width=\linewidth]{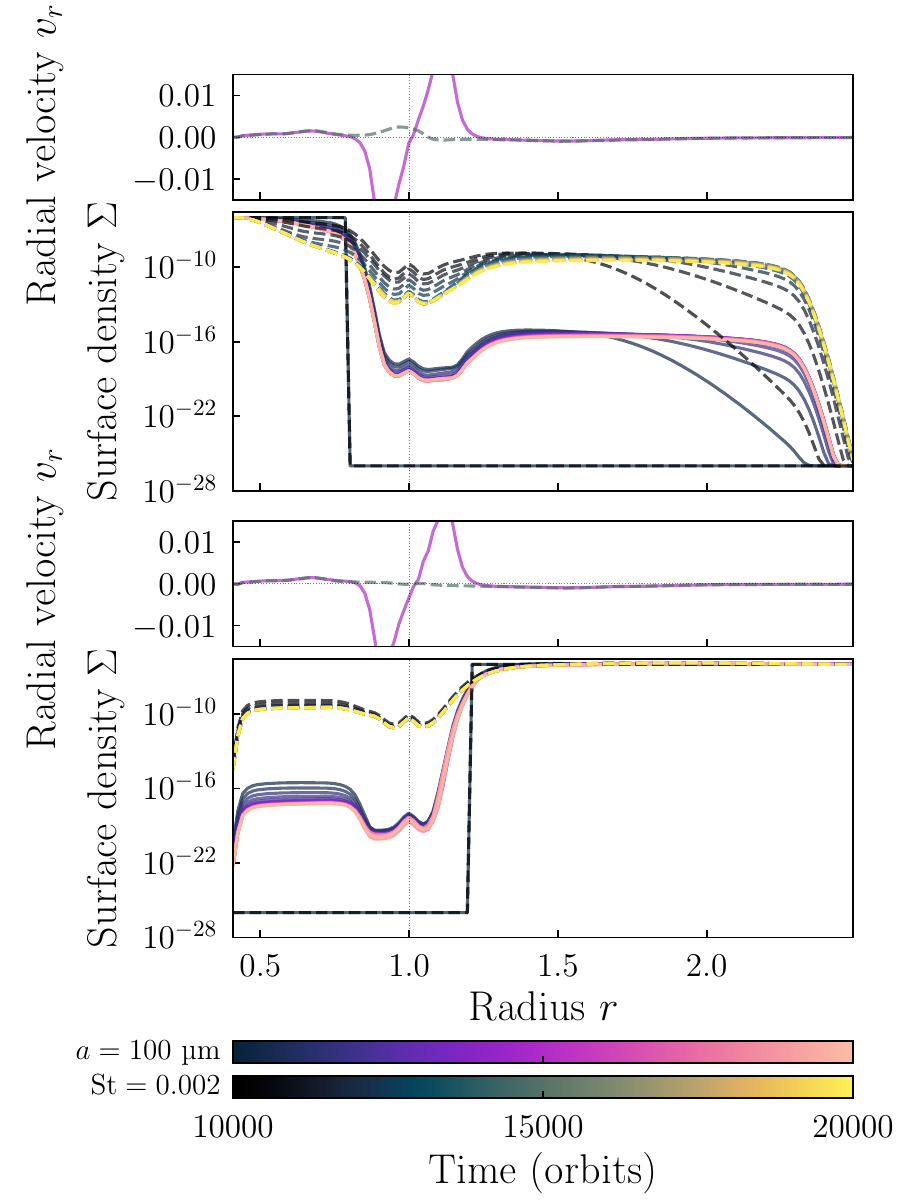}
    \caption{Same as Figure~\ref{fig:stepfn-30earth-100um}, but for a disk containing a 1~\MJup\ planet.}
    \label{fig:stepfn-jupiter-100um}
\end{figure}

\begin{figure}
    \centering
    \includegraphics[width=\linewidth]{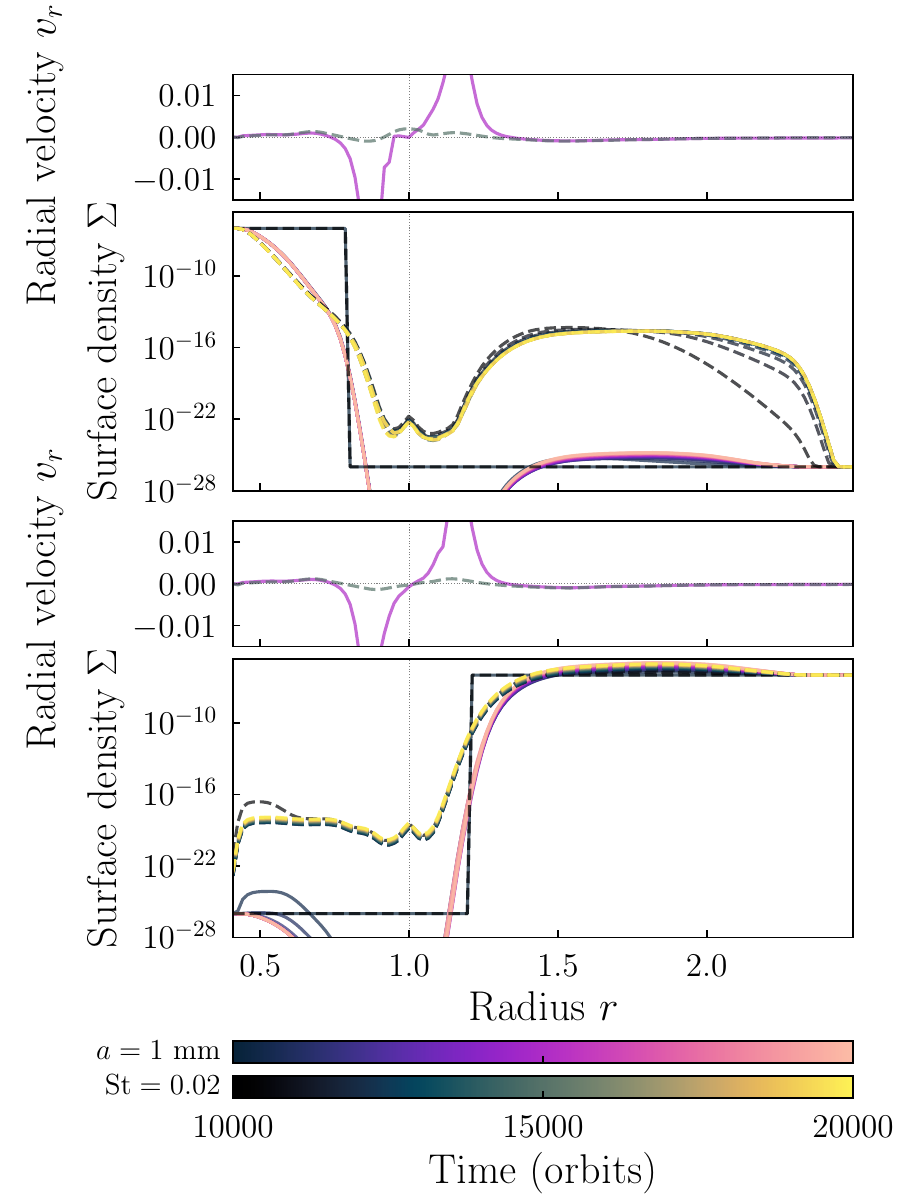}
    \caption{Same as Figure~\ref{fig:stepfn-jupiter-100um}, but for $a = 1$~\mm\ (solid lines) and $\Stokes = 2 \times 10^{-2}$ (dashed lines).}
    \label{fig:stepfn-jupiter-1mm}
\end{figure}

To better understand the inward and outward migration of solids across the orbit of the planet, we follow \citet{Weber+2018} in evolving a step function dust distribution from interior to or exterior to the planet's orbit, well after the gas surface density has been sculpted by the planet. In FARGO3D, we accomplish this by disabling all but orbital motion on the dust fluids until $10^4$ orbits; then, we enable the remaining forces and evolve the entire system for an additional $10^4$ orbits. Azimuthally-averaged surface density profiles that result from evolving an initial step function profile are shown in Figures~\ref{fig:stepfn-30earth-100um}, \ref{fig:stepfn-30earth-1mm}, \ref{fig:stepfn-jupiter-100um}, and \ref{fig:stepfn-jupiter-1mm}.

In Figure~\ref{fig:stepfn-30earth-100um}, we show the surface density evolution in the presence of a 30~\MEarth\ planet for particles with size $a = 100$~\um\ and $\Stokes = 0.002$. After $2 \times 10^4$ orbits, for dust that starts interior to the planet's orbit, the surface density of the fixed size particles is higher outside the planet's orbit than that of fixed Stokes number particles, by about an order of magnitude. Since the fixed Stokes number dust has a higher Stokes number than the fixed size dust in the inner disk (see Figure~\ref{fig:stokes-map-jupiter}), the fixed Stokes number dust is more likely to drift inward than be caught in the accretion flow past the planet and into the outer disk. In the outer disk, the expanding dust density front extends to about the same orbital radius at a given time regardless of the aerodynamics assumptions. As shown in Figure~\ref{fig:stokes-map-jupiter}, the constant Stokes number approximation is more accurate in the outer disk, so we expect the dynamics in the outer disk to be very similar between constant size and constant Stokes number particles.

When the $a = 100$~\um\ and $\Stokes = 0.002$ dust starts exterior to the planet's orbit, there is significant inward drift of both the fixed size and fixed Stokes number dusts, achieving a surface density in the inner disk about $25\%$ of that in the outer disk. The 30~\MEarth\ planet is an ineffective barrier to inward drift of both kinds of material, and, because fixed size material in the outer disk has roughly the same Stokes number as the fixed Stokes material, the dynamics of dust crossing the planet into the inner disk are very similar.

Figure~\ref{fig:stepfn-30earth-1mm} shows the evolution of the dust surface density in the same simulations, but for the $a = 1$~\mm\ and $\Stokes = 0.02$ dust fluids. There is significantly less outward drift from the inner disk at this size, but the constant size particles again experience more outward drift than the constant Stokes number particles. The surface density of constant size $a = 1$~\mm\ particles in the outer disk exceeds that of the constant Stokes number $\Stokes = 0.02$ particles by orders of magnitude, and the effect becomes more pronounced over time, suggesting ongoing outward motion. The planet-induced gap is additionally much deeper and wider when Stokes number is fixed instead of particle size. The Stokes number of the fixed size dust in the inner disk is lower than the corresponding fixed value (see Figure~\ref{fig:stokes-map-jupiter}), so the fixed size dust is better coupled to the gas in the inner disk. The fixed Stokes number dust with $\Stokes = 0.02$ will drift inward more rapidly with no barrier (as seen from its radial velocity in Figure~\ref{fig:stepfn-30earth-100um}), draining the inner disk reservoir. The fixed size dust with $a = 1$~\mm, on the other hand, is better entrained in the gas and can be carried past the planet more efficiently when it is caught up in the accretion flow.

When the $a = 1$~\mm\ and $\Stokes = 0.02$ dust is initialized in the outer disk, we still observe some flux inward across the planet-induced gap, albeit less than in the smaller dust size discussed above, by about an order of magnitude. There is slightly more fixed size dust in the inner disk than fixed Stokes number dust after $2 \times 10^4$ orbits, though the surface density bump immediately exterior to the planet's orbit is roughly the same magnitude independent of the assumptions on aerodynamics. In the outer disk, the fixed size dust has roughly the same Stokes number as the fixed Stokes number dust, and so we confirm that the dynamics there are very similar. Fixed size dust that does cross to the inner disk has a lower Stokes number than the corresponding constant value, so the fixed Stokes number dust should be depleted more efficiently, resulting in the observed lower surface density of $\Stokes = 0.02$ dust in the inner disk.

Next, we increase the planet mass to 1~\MJup\ and perform the same analysis, finding notable differences in the sculpting of the step function dust profile. Figure~\ref{fig:stepfn-jupiter-100um} is analogous to Figure~\ref{fig:stepfn-30earth-100um} but corresponds to the more massive planet. With the mass increase, we expect that the gap in the dust surface density becomes more prominent (lower density and wider), and indeed this holds in general. When the $a = 100$~\um\ or $\Stokes = 0.002$ dust all originates from the inner disk, we see that the inner gap edge is sharper when size is fixed rather than Stokes number. Significantly more dust (a factor of more than five orders of magnitude in surface density) crosses the planet-induced gap outward when Stokes number is fixed rather than size. Fixed size $a = 100$~\um\ dust that approaches the 1~\MJup\ planet has significantly larger Stokes number (see Figure~\ref{fig:stokes-map-jupiter}), so it becomes more decoupled from the gas flow and may begin to drift back towards the central star; this negative feedback can explain the lack of outward drift for the fixed size $a = 100$~\um\ dust compared to its $\Stokes = 0.002$ counterpart.

When dust is seeded in the outer disk instead, we again observe that a large amount of the fixed Stokes number material crosses to the inner disk, where it is slowly depleted by drift over time, but almost none of the fixed size material crosses the gap. As is the case for the 30~\MEarth\ planet, the approximation of Stokes number in the outer disk is much better than in the inner disk, and so we expect dynamics in the outer disk to be similar regardless of whether the particle size or Stokes number is fixed. Dust of fixed size $a = 100$~\um\ that does approach the planet has a larger Stokes number than its fixed Stokes number counterpart, so the $\Stokes = 0.002$ dust is carried more easily past the planet by the gas streams.

Finally, we examine the effect of a 1~\MJup\ planet on step function surface density profiles of $a = 1$~\mm\ and $\Stokes = 0.02$ dust populations in Figure~\ref{fig:stepfn-jupiter-1mm}. For a step function seeded in the inner disk, the behavior noted for the $a = 100~\um$ and $\Stokes = 0.002$ dust still applies, though less of the fixed Stokes number material crosses to the outer disk than before, and almost no fixed size material crosses the gap. These grains are more affected by drift due to gas drag, so they are less likely to migrate outwards. When the step function is initialized in the outer disk, there is some inward migration of the fixed Stokes number solids, but the surface density in the inner disk remains more than ten orders of magnitude lower than that in the outer disk, indicating a much lower efficiency of crossing than for smaller particles. This behavior is generally consistent with previous studies \citep[among others]{Kruijer+2017} which found that a Jupiter-mass planet should be an effective barrier to large solids.

We do not show analogous figures for the smaller dust sizes ($a < 100~\um$) and Stokes numbers ($\Stokes < 2 \times 10^{-3}$) at either planet mass. In the case of the 30~\MEarth\ planet, at dust sizes $a \lesssim 100~\um$ and Stokes numbers $\Stokes \lesssim 2 \times 10^{-3}$, we observe more inward migration than outward migration, as expected, with slightly more material crossing the planet-induced gap outward when size is fixed rather than Stokes number. For the larger, 1~\MJup\ planet, more small material crosses the gap in both directions when the Stokes number is fixed. There is a clear transition to more dramatic differences between the two drag treatments at $a \sim 100~\um$, which is justified by the discrepancy in the measured and assumed Stokes number shown in Figure~\ref{fig:stokes-map-jupiter} and the nonlinear dependence of dust stopping time on the Stokes number. The 100~\um\ dust is large enough that the increase in Stokes number has a significant effect on dynamics.

In addition to comparing large-scale features in the surface densities and evolution of a dust step function, we can make detailed comparisons of the simulated particle trajectories, which provide more information about the conditions an individual particle would experience as it moves through the disk. We did not find this information to be as illuminating as the results above, however, possibly due to simulating too few trajectories. As we show, tens of thousands of trajectories might be needed to observe a single particle successfully cross a planet-induced gap. We do observe some variation in the fraction of particles that cross the gap (in both directions) with particle size and release time.

\section{Discussion}
\label{sec:discussion}

\subsection{Implications for the CAI storage problem}

\noindent The presence of CAIs (calcium-rich, aluminum-rich inclusions) in chondritic meteorites has long presented a challenge for solar nebula dynamics, as these objects should have formed in a high-temperature environment \citep[e.g.][]{GrossmanLarimer1974}, near the young Sun.  However, most CAIs are found in CC meteorites, which are believed to have formed far from the Sun, beyond the water snow line , and outside of Jupiter’s orbit \citep[e.g.,][]{Kruijer+2020}. CAIs are much less abundant, and generally smaller, in NC meteorites \citet{Dunham+2023}, which formed closer to the Sun.

\citet{Desch+2018} develop a model of the protosolar disk, informed by measurements of meteorite samples, to address the so-called ``CAI storage problem,'' outlined above. Their model includes the formation of CAIs from the parent refractory elements, concentrated by turbulence and incorporated into the final meteorite bodies. \citet{Desch+2018} find that the relative abundance of CAIs, defined as objects with radii of 2500~\um, compared to all solids, is enhanced in the outer disk because the CAI material cannot cross the gap in the protosolar disk carved out by a forming Jupiter; in the inner disk, the CAI material is depleted because it drifts rapidly into the young Sun.

Our findings also suggest millimeter-sized objects and larger would get preserved outside the planet’s orbit, though we find that the dynamics of smaller grains require consideration. In Section~\ref{sec:stepfn}, we found that the 100~\um\ particles frequently have nonzero gap-crossing fluxes even when 1~\mm\ particles are blocked almost entirely. The 100~\um\ size bin is particularly interesting in the context of CAIs because the size distribution of CAIs measured in meteorite samples peaks around $\sim 100~\um$, including those found in NC meteorites \citep{Simon2018,Simon2018corr,Dunham+2023}. We find that both the inward and outward flux of those particles depends not only on the planet mass and particle release time, as expected, but also on the assumption of constant particle size \textit{versus} constant Stokes number. This again highlights the need for a more accurate treatment of particle dynamics using the physical measure of particle size over Stokes number, since the Stokes number depends on local gas conditions.

While the nonzero flux of 100~\um\ particles crossing the planet-induced gap outwards provides a means for CAIs that formed close to the Sun to be incorporated into meteoritic material in the outer disk, most of the particles do not drift outward and would actually have been incorporated into inner Solar System material, which is inconsistent with the observation of very few CAIs in NC meteorites. Overall, our simulations show that outward transport within the disk, while nonzero, is inefficient, suggesting that, if it occurs, it peaks before large planets carve gaps in their protoplanetary disks.  Thus, if transported within the disk CAIs must have been delivered from the inner regions to outside Jupiter’s orbit very early in Solar System history, prior to it reaching to $\sim 10~\MEarth$; this is consistent with early stages of disks having high rates of mass and angular momentum transport and the CAIs being the oldest objects in the Solar System \citep{Ciesla2010}.

We note, however, that recent studies \citep{Schrader+2020,Schrader+2022} do find isotopic signatures in meteorite samples that suggest outward migration of rocky material later in solar nebula history. Here we show such transport can happen, even with a sufficiently massive young Jupiter present.

\subsection{Limitations of this study}

We have argued above that one-dimensional simulations of protoplanetary disks with an embedded planet are inherently limited because they cannot capture features such as spiral waves, which can only be observed in two dimensions. A similar criticism can be made for simulations in two dimensions, however, which cannot capture three-dimensional phenomena like meridional flows and dust settling, which may have a significant effect on simulated observations of disks \citep[e.g.,][]{Dipierro+2015}. We refer the reader to future work, with simulations of constant size particles using FARGO3D in three dimensions \citep{VanClepper+2025}.

\section{Conclusions}
\label{sec:conclusions}

\noindent We have found that both inward and outward drift of solids is observed in our models independently of whether the particle size or Stokes number is fixed, but that the specifics of disk morphology and particle flux across a planet-induced gap do change depending on how the dynamics are computed. Inward transport past a forming planet can be efficient before the gap is fully formed, especially for particles smaller than 1~\mm. To carry out our simulations, we derive, in Appendix~\ref{adx:analytic}, a novel set of analytic equations for particle dynamics at fixed size that can be incorporated into any numerical hydrodynamics code that employs operator splitting techniques.

\section*{Acknowledgements}
    EMP thanks the Heising-Simons Foundation for their generous support through the 51 Pegasi b postdoctoral fellowship. Additionally, EMP acknowledges helpful discussion with Dr. Leonardo Krapp (Universidad de Concepci\'on), particularly during peer review. EVC acknowledges support from NASA FINESST grant 80NSSC23K1380. This work was completed in part with resources provided by the University of Chicago Research Computing Center.

    This material is based upon work supported by the National Aeronautics and Space Administration under Agreement No. 80NSSC21K0593 for the program ``Alien Earths''. The results reported herein benefited from collaborations and/or information exchange within NASA’s Nexus for Exoplanet System Science (NExSS) research coordination network sponsored by NASA’s Science Mission Directorate.

\software{FARGO3D \citep{fargo3dsoftware}}

\bibliographystyle{aasjournal}
\bibliography{biblio}

\begin{thebibliography}{}
\expandafter\ifx\csname natexlab\endcsname\relax\def\natexlab#1{#1}\fi
\providecommand{\url}[1]{\href{#1}{#1}}
\providecommand{\dodoi}[1]{doi:~\href{http://doi.org/#1}{\nolinkurl{#1}}}
\providecommand{\doeprint}[1]{\href{http://ascl.net/#1}{\nolinkurl{http://ascl.net/#1}}}
\providecommand{\doarXiv}[1]{\href{https://arxiv.org/abs/#1}{\nolinkurl{https://arxiv.org/abs/#1}}}

\bibitem[{{Auddy} {et~al.}(2022){Auddy}, {Dey}, {Lin}, {Carrera}, \&
  {Simon}}]{Auddy+2022}
{Auddy}, S., {Dey}, R., {Lin}, M.-K., {Carrera}, D., \& {Simon}, J.~B. 2022,
  \apj, 936, 93, \dodoi{10.3847/1538-4357/ac7a3c}

\bibitem[{{Ayliffe} {et~al.}(2012){Ayliffe}, {Laibe}, {Price}, \&
  {Bate}}]{Ayliffe+2012}
{Ayliffe}, B.~A., {Laibe}, G., {Price}, D.~J., \& {Bate}, M.~R. 2012, \mnras,
  423, 1450, \dodoi{10.1111/j.1365-2966.2012.20967.x}

\bibitem[{{Ben{\'\i}tez-Llambay} {et~al.}(2019){Ben{\'\i}tez-Llambay}, {Krapp},
  \& {Pessah}}]{fargo3dmultifluid}
{Ben{\'\i}tez-Llambay}, P., {Krapp}, L., \& {Pessah}, M.~E. 2019, \apjs, 241,
  25, \dodoi{10.3847/1538-4365/ab0a0e}

\bibitem[{{Ben{\'\i}tez Llambay} \& {Masset}(2015)}]{fargo3dsoftware}
{Ben{\'\i}tez Llambay}, P., \& {Masset}, F. 2015, {FARGO3D:
  Hydrodynamics/magnetohydrodynamics code}, Astrophysics Source Code Library,
  record ascl:1509.006.
\newblock \doeprint{1509.006}

\bibitem[{{Ben{\'\i}tez-Llambay} \& {Masset}(2016)}]{fargo3d}
{Ben{\'\i}tez-Llambay}, P., \& {Masset}, F.~S. 2016, \apjs, 223, 11,
  \dodoi{10.3847/0067-0049/223/1/11}

\bibitem[{{Bermingham} {et~al.}(2018){Bermingham}, {Worsham}, \&
  {Walker}}]{Bermingham+2018}
{Bermingham}, K.~R., {Worsham}, E.~A., \& {Walker}, R.~J. 2018, Earth and
  Planetary Science Letters, 487, 221, \dodoi{10.1016/j.epsl.2018.01.017}

\bibitem[{{Binkert} {et~al.}(2021){Binkert}, {Szul{\'a}gyi}, \&
  {Birnstiel}}]{Binkert+2021}
{Binkert}, F., {Szul{\'a}gyi}, J., \& {Birnstiel}, T. 2021, \mnras, 506, 5969,
  \dodoi{10.1093/mnras/stab2075}

\bibitem[{{Binkert} {et~al.}(2023){Binkert}, {Szul{\'a}gyi}, \&
  {Birnstiel}}]{Binkert+2023}
---. 2023, \mnras, 523, 55, \dodoi{10.1093/mnras/stad1405}

\bibitem[{{Birnstiel} {et~al.}(2010){Birnstiel}, {Dullemond}, \&
  {Brauer}}]{Birnstiel+2010}
{Birnstiel}, T., {Dullemond}, C.~P., \& {Brauer}, F. 2010, \aap, 513, A79,
  \dodoi{10.1051/0004-6361/200913731}

\bibitem[{{Budde} {et~al.}(2016){Budde}, {Burkhardt}, {Brennecka},
  {Fischer-G{\"o}dde}, {Kruijer}, \& {Kleine}}]{Budde+2016}
{Budde}, G., {Burkhardt}, C., {Brennecka}, G.~A., {et~al.} 2016, Earth and
  Planetary Science Letters, 454, 293, \dodoi{10.1016/j.epsl.2016.09.020}

\bibitem[{{Budde} {et~al.}(2018){Budde}, {Kruijer}, \& {Kleine}}]{Budde+2018}
{Budde}, G., {Kruijer}, T.~S., \& {Kleine}, T. 2018, \gca, 222, 284,
  \dodoi{10.1016/j.gca.2017.10.014}

\bibitem[{{Chan} \& {Paardekooper}(2024)}]{Chan+2024}
{Chan}, K., \& {Paardekooper}, S.-J. 2024, \mnras, 528, 5904,
  \dodoi{10.1093/mnras/stae089}

\bibitem[{{Ciesla}(2010)}]{Ciesla2010}
{Ciesla}, F.~J. 2010, \icarus, 208, 455, \dodoi{10.1016/j.icarus.2010.02.010}

\bibitem[{{Desch} {et~al.}(2018){Desch}, {Kalyaan}, \& {O'D.
  Alexander}}]{Desch+2018}
{Desch}, S.~J., {Kalyaan}, A., \& {O'D. Alexander}, C.~M. 2018, \apjs, 238, 11,
  \dodoi{10.3847/1538-4365/aad95f}

\bibitem[{{Dipierro} {et~al.}(2015){Dipierro}, {Price}, {Laibe}, {Hirsh},
  {Cerioli}, \& {Lodato}}]{Dipierro+2015}
{Dipierro}, G., {Price}, D., {Laibe}, G., {et~al.} 2015, \mnras, 453, L73,
  \dodoi{10.1093/mnrasl/slv105}

\bibitem[{{Dr{\k{a}}{\.z}kowska} {et~al.}(2019){Dr{\k{a}}{\.z}kowska}, {Li},
  {Birnstiel}, {Stammler}, \& {Li}}]{Drazkowska+2019}
{Dr{\k{a}}{\.z}kowska}, J., {Li}, S., {Birnstiel}, T., {Stammler}, S.~M., \&
  {Li}, H. 2019, \apj, 885, 91, \dodoi{10.3847/1538-4357/ab46b7}

\bibitem[{{Dullemond} {et~al.}(2022){Dullemond}, {Ziampras}, {Ostertag}, \&
  {Dominik}}]{Dullemond+2022}
{Dullemond}, C.~P., {Ziampras}, A., {Ostertag}, D., \& {Dominik}, C. 2022,
  \aap, 668, A105, \dodoi{10.1051/0004-6361/202244218}

\bibitem[{{Dunham} {et~al.}(2023){Dunham}, {Sheikh}, {Opara}, {Matsuda}, {Liu},
  \& {McKeegan}}]{Dunham+2023}
{Dunham}, E.~T., {Sheikh}, A., {Opara}, D., {et~al.} 2023, \maps, 58, 643,
  \dodoi{10.1111/maps.13975}

\bibitem[{{Eriksson} {et~al.}(2020){Eriksson}, {Johansen}, \&
  {Liu}}]{Eriksson+2020}
{Eriksson}, L. E.~J., {Johansen}, A., \& {Liu}, B. 2020, \aap, 635, A110,
  \dodoi{10.1051/0004-6361/201937037}

\bibitem[{{Fouchet} {et~al.}(2007){Fouchet}, {Maddison}, {Gonzalez}, \&
  {Murray}}]{Fouchet+2007}
{Fouchet}, L., {Maddison}, S.~T., {Gonzalez}, J.~F., \& {Murray}, J.~R. 2007,
  \aap, 474, 1037, \dodoi{10.1051/0004-6361:20077586}

\bibitem[{{F{\"u}ri} \& {Marty}(2015)}]{FuriMarty2015}
{F{\"u}ri}, E., \& {Marty}, B. 2015, Nature Geoscience, 8, 515,
  \dodoi{10.1038/ngeo2451}

\bibitem[{{Grossman} \& {Larimer}(1974)}]{GrossmanLarimer1974}
{Grossman}, L., \& {Larimer}, J.~W. 1974, Reviews of Geophysics and Space
  Physics, 12, 71, \dodoi{10.1029/RG012i001p00071}

\bibitem[{{Huang} \& {Bai}(2022)}]{Huang+2022}
{Huang}, P., \& {Bai}, X.-N. 2022, \apjs, 262, 11,
  \dodoi{10.3847/1538-4365/ac76cb}

\bibitem[{{Kruijer} {et~al.}(2017){Kruijer}, {Burkhardt}, {Budde}, \&
  {Kleine}}]{Kruijer+2017}
{Kruijer}, T.~S., {Burkhardt}, C., {Budde}, G., \& {Kleine}, T. 2017,
  Proceedings of the National Academy of Science, 114, 6712,
  \dodoi{10.1073/pnas.1704461114}

\bibitem[{{Kruijer} {et~al.}(2020){Kruijer}, {Kleine}, \&
  {Borg}}]{Kruijer+2020}
{Kruijer}, T.~S., {Kleine}, T., \& {Borg}, L.~E. 2020, Nature Astronomy, 4, 32,
  \dodoi{10.1038/s41550-019-0959-9}

\bibitem[{{Laibe} \& {Price}(2012)}]{LaibePrice2012}
{Laibe}, G., \& {Price}, D.~J. 2012, \mnras, 420, 2365,
  \dodoi{10.1111/j.1365-2966.2011.20201.x}

\bibitem[{{Lin} \& {Papaloizou}(1986)}]{LinPaploizou1986}
{Lin}, D.~N.~C., \& {Papaloizou}, J. 1986, \apj, 309, 846,
  \dodoi{10.1086/164653}

\bibitem[{{Liu} {et~al.}(2022){Liu}, {Johansen}, {Lambrechts}, {Bizzarro}, \&
  {Haugb{\o}lle}}]{Liu+2022}
{Liu}, B., {Johansen}, A., {Lambrechts}, M., {Bizzarro}, M., \& {Haugb{\o}lle},
  T. 2022, Science Advances, 8, eabm3045, \dodoi{10.1126/sciadv.abm3045}

\bibitem[{{Lyra} {et~al.}(2009){Lyra}, {Johansen}, {Klahr}, \&
  {Piskunov}}]{Lyra+2009}
{Lyra}, W., {Johansen}, A., {Klahr}, H., \& {Piskunov}, N. 2009, \aap, 493,
  1125, \dodoi{10.1051/0004-6361:200810797}

\bibitem[{{Maddison} {et~al.}(2007){Maddison}, {Fouchet}, \&
  {Gonzalez}}]{Maddison+2007}
{Maddison}, S.~T., {Fouchet}, L., \& {Gonzalez}, J.~F. 2007, \apss, 311, 3,
  \dodoi{10.1007/s10509-007-9572-y}

\bibitem[{{Mignone} {et~al.}(2007){Mignone}, {Bodo}, {Massaglia}, {Matsakos},
  {Tesileanu}, {Zanni}, \& {Ferrari}}]{pluto}
{Mignone}, A., {Bodo}, G., {Massaglia}, S., {et~al.} 2007, \apjs, 170, 228,
  \dodoi{10.1086/513316}

\bibitem[{{Mignone} {et~al.}(2019){Mignone}, {Flock}, \&
  {Vaidya}}]{Mignone+2019}
{Mignone}, A., {Flock}, M., \& {Vaidya}, B. 2019, \apjs, 244, 38,
  \dodoi{10.3847/1538-4365/ab4356}

\bibitem[{{Moseley} {et~al.}(2023){Moseley}, {Teyssier}, \&
  {Draine}}]{Moseley+2023}
{Moseley}, E.~R., {Teyssier}, R., \& {Draine}, B.~T. 2023, \mnras, 518, 2825,
  \dodoi{10.1093/mnras/stac3231}

\bibitem[{{Moynier} {et~al.}(2012){Moynier}, {Day}, {Okui}, {Yokoyama},
  {Bouvier}, {Walker}, \& {Podosek}}]{Moynier+2012}
{Moynier}, F., {Day}, J. M.~D., {Okui}, W., {et~al.} 2012, \apj, 758, 45,
  \dodoi{10.1088/0004-637X/758/1/45}

\bibitem[{{Nanne} {et~al.}(2019){Nanne}, {Nimmo}, {Cuzzi}, \&
  {Kleine}}]{Nanne+2019}
{Nanne}, J. A.~M., {Nimmo}, F., {Cuzzi}, J.~N., \& {Kleine}, T. 2019, Earth and
  Planetary Science Letters, 511, 44, \dodoi{10.1016/j.epsl.2019.01.027}

\bibitem[{{Paardekooper}(2007)}]{Paardekooper2007}
{Paardekooper}, S.~J. 2007, \aap, 462, 355, \dodoi{10.1051/0004-6361:20066326}

\bibitem[{{Paardekooper} \& {Mellema}(2004)}]{Paardekooper+2004}
{Paardekooper}, S.~J., \& {Mellema}, G. 2004, \aap, 425, L9,
  \dodoi{10.1051/0004-6361:200400053}

\bibitem[{{Paardekooper} \& {Mellema}(2006)}]{Paardekooper+2006}
---. 2006, \aap, 453, 1129, \dodoi{10.1051/0004-6361:20054449}

\bibitem[{{Pierens} {et~al.}(2019){Pierens}, {Lin}, \&
  {Raymond}}]{Pierens+2019}
{Pierens}, A., {Lin}, M.~K., \& {Raymond}, S.~N. 2019, \mnras, 488, 645,
  \dodoi{10.1093/mnras/stz1718}

\bibitem[{{Price} {et~al.}(2018){Price}, {Wurster}, {Tricco}, {Nixon},
  {Toupin}, {Pettitt}, {Chan}, {Mentiplay}, {Laibe}, {Glover}, {Dobbs},
  {Nealon}, {Liptai}, {Worpel}, {Bonnerot}, {Dipierro}, {Ballabio}, {Ragusa},
  {Federrath}, {Iaconi}, {Reichardt}, {Forgan}, {Hutchison}, {Constantino},
  {Ayliffe}, {Hirsh}, \& {Lodato}}]{Price+2018}
{Price}, D.~J., {Wurster}, J., {Tricco}, T.~S., {et~al.} 2018, \pasa, 35, e031,
  \dodoi{10.1017/pasa.2018.25}

\bibitem[{{Schrader} \& {Davidson}(2022)}]{Schrader+2022}
{Schrader}, D.~L., \& {Davidson}, J. 2022, Earth and Planetary Science Letters,
  589, 117552, \dodoi{10.1016/j.epsl.2022.117552}

\bibitem[{{Schrader} {et~al.}(2020){Schrader}, {Nagashima}, {Davidson},
  {McCoy}, {Ogliore}, \& {Fu}}]{Schrader+2020}
{Schrader}, D.~L., {Nagashima}, K., {Davidson}, J., {et~al.} 2020, \gca, 282,
  133, \dodoi{10.1016/j.gca.2020.05.014}

\bibitem[{{Shakura} \& {Sunyaev}(1973)}]{ShakuraSunyaev1973}
{Shakura}, N.~I., \& {Sunyaev}, R.~A. 1973, \aap, 24, 337

\bibitem[{{Simon} {et~al.}(2018{\natexlab{a}}){Simon}, {Cuzzi}, {McCain},
  {Cato}, {Christoffersen}, {Fisher}, {Srinivasan}, {Tait}, {Olson}, \&
  {Scargle}}]{Simon2018}
{Simon}, J.~I., {Cuzzi}, J.~N., {McCain}, K.~A., {et~al.} 2018{\natexlab{a}},
  Earth and Planetary Science Letters, 494, 69,
  \dodoi{10.1016/j.epsl.2018.04.021}

\bibitem[{{Simon} {et~al.}(2018{\natexlab{b}}){Simon}, {Cuzzi}, {McCain},
  {Cato}, {Christoffersen}, {Fisher}, {Srinivasan}, {Tait}, {Olson}, \&
  {Scargle}}]{Simon2018corr}
---. 2018{\natexlab{b}}, Earth and Planetary Science Letters, 502, 293,
  \dodoi{10.1016/j.epsl.2018.08.022}

\bibitem[{{Stammler} {et~al.}(2023){Stammler}, {Lichtenberg},
  {Dr{\k{a}}{\.z}kowska}, \& {Birnstiel}}]{Stammler+2023}
{Stammler}, S.~M., {Lichtenberg}, T., {Dr{\k{a}}{\.z}kowska}, J., \&
  {Birnstiel}, T. 2023, \aap, 670, L5, \dodoi{10.1051/0004-6361/202245512}

\bibitem[{{Sturm} {et~al.}(2020){Sturm}, {Rosotti}, \& {Dominik}}]{Sturm+2020}
{Sturm}, J.~A., {Rosotti}, G.~P., \& {Dominik}, C. 2020, \aap, 643, A92,
  \dodoi{10.1051/0004-6361/202038919}

\bibitem[{{Van Clepper} {et~al.}(2025){Van Clepper}, {Price}, \&
  {Ciesla}}]{VanClepper+2025}
{Van Clepper}, E., {Price}, E.~M., \& {Ciesla}, F.~J. 2025, arXiv e-prints,
  arXiv:2501.07520, \dodoi{10.48550/arXiv.2501.07520}

\bibitem[{{Warren}(2011)}]{Warren2011}
{Warren}, P.~H. 2011, Earth and Planetary Science Letters, 311, 93,
  \dodoi{10.1016/j.epsl.2011.08.047}

\bibitem[{{Weber} {et~al.}(2018){Weber}, {Ben{\'\i}tez-Llambay}, {Gressel},
  {Krapp}, \& {Pessah}}]{Weber+2018}
{Weber}, P., {Ben{\'\i}tez-Llambay}, P., {Gressel}, O., {Krapp}, L., \&
  {Pessah}, M.~E. 2018, \apj, 854, 153, \dodoi{10.3847/1538-4357/aaab63}

\bibitem[{{Wu} {et~al.}(2023){Wu}, {Baruteau}, \& {Nayakshin}}]{Wu+2023}
{Wu}, Y., {Baruteau}, C., \& {Nayakshin}, S. 2023, \mnras, 523, 4869,
  \dodoi{10.1093/mnras/stad1791}

\bibitem[{{Yokoyama} {et~al.}(2015){Yokoyama}, {Fukami}, {Okui}, {Ito}, \&
  {Yamazaki}}]{Yokoyama+2015}
{Yokoyama}, T., {Fukami}, Y., {Okui}, W., {Ito}, N., \& {Yamazaki}, H. 2015,
  Earth and Planetary Science Letters, 416, 46,
  \dodoi{10.1016/j.epsl.2015.01.040}

\end{thebibliography}

\appendix

\section{Analytic drag formulae}
\label{adx:analytic}
\restartappendixnumbering

\begin{figure*}
    \centering
    \includegraphics[width=\linewidth]{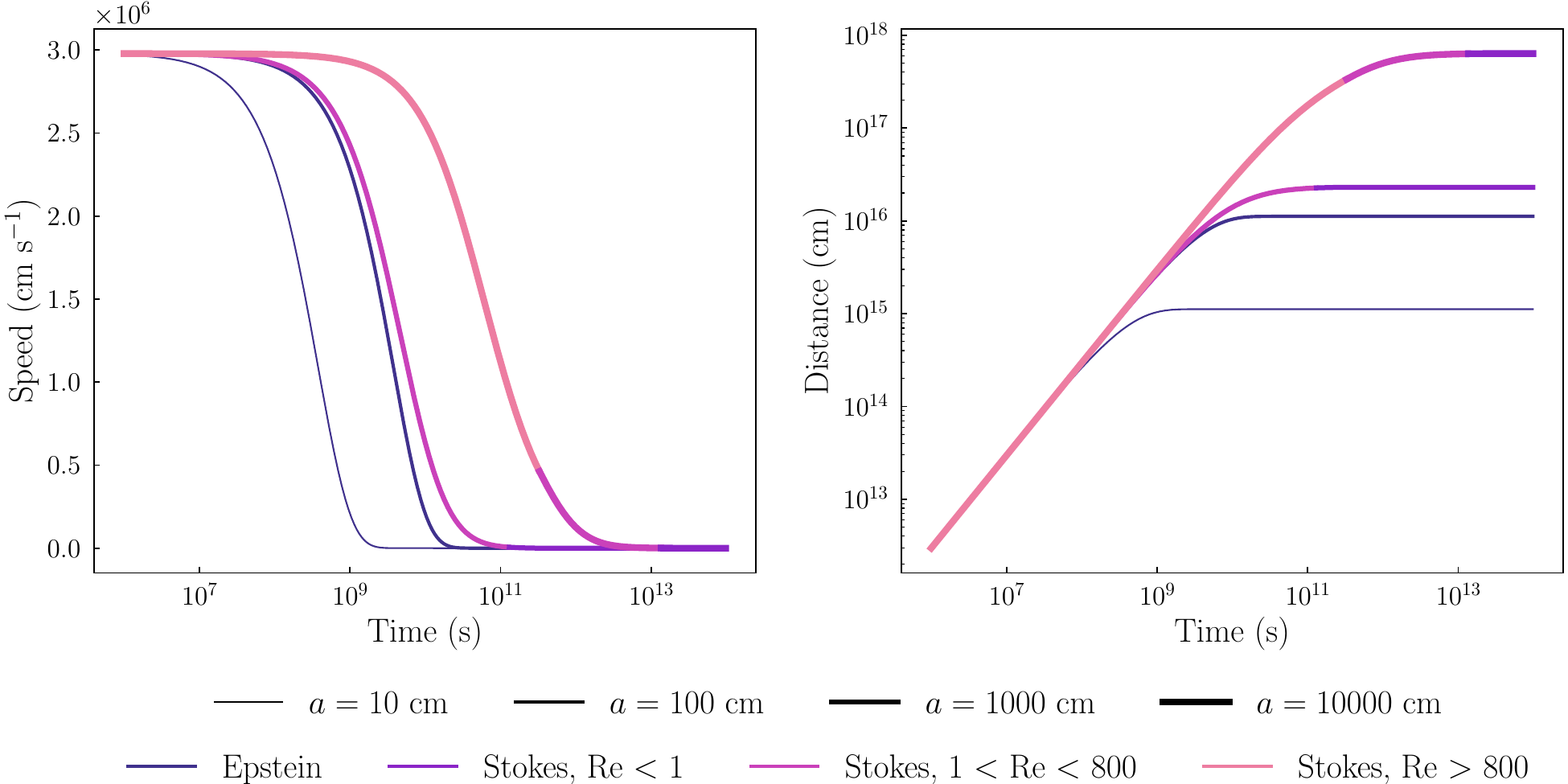}
    \caption{Overview of possible analytic solutions computed using the formulae in Appendix~\ref{adx:analytic}. Four different particle sizes are shown, starting from rest in a dilute gas moving at Keplerian velocity in a toy model. Holding the gas conditions fixed, the particle speed and distance are computed analytically through all possible regimes.}
    \label{fig:analytic-solution-zoo}
\end{figure*}

\noindent For the purposes of this derivation, the Reynolds number is given by
\begin{equation}
    \Reynolds = \frac{2 a \lVert \Delta \bm{u} \rVert}{\nu}.
\end{equation}
The vector $\Delta \bm{u} = \bm{u}_\mathrm{dust} - \bm{u}_\mathrm{gas}$ is the velocity difference between the gas and dust; $m$ is the mass of a single dust particle; $\nu$ is the gas viscosity; $a$ is the particle radius; and $\rho = \rho_\mathrm{gas} + \rho_\mathrm{dust}$ is the total volume density of the gas and dust species under consideration. For the differential equations that apply in various drag regimes stated in this derivation, we refer the reader to \citet{LaibePrice2012} and references therein. We provide Figure~\ref{fig:analytic-solution-zoo} to illustrate possible analytic solutions to the equations derived below, across the Epstein and Stokes regimes and for a few different particle sizes.

\subsection{Stokes regime}

\noindent Within the Stokes regime, there are three sub-regimes we must consider to treat drag correctly, since the drag coefficient $\dragcoeff$ is a piecewise function of Reynolds number. In the Stokes regime, the rate of change of the velocity difference, is given by
\begin{equation}
    \partial_t \left(\Delta \bm{u}\right) = -\frac{\pi a^2 \rho}{2 m} \dragcoeff \lVert \Delta \bm{u} \rVert \Delta \bm{u}.
\end{equation}

\subsubsection{Small Reynolds number}

\noindent When $\Reynolds < 1$, we have $\dragcoeff = 24 \Reynolds^{-1}$, so
\begin{equation}
    \partial_t \left(\Delta \bm{u}\right) = -\frac{6 \pi a \nu \rho}{m} \Delta \bm{u}.
\end{equation}
This equation is linear because the factors of $\lVert \Delta \bm{u} \rVert$ cancel, so the solution is trivially
\begin{equation}
    \Delta \bm{u}\!\left(t\right) = \exp\!\left(-\frac{6\pi a \nu \rho t}{m}\right) \Delta \bm{u}_0.
    \label{eqn:stokessmall}
\end{equation}

\subsubsection{Intermediate Reynolds number}

\noindent For $1 < \Reynolds < 800$, we have $\dragcoeff = 24 \Reynolds^{-3/5}$, so
\begin{equation}
    \partial_t \left(\Delta \bm{u}\right) = -\frac{6 \pi a \nu \rho}{m} \left(\frac{2 a}{\nu}\right)^{2/5} \lVert \Delta \bm{u} \rVert^{2/5} \Delta \bm{u}.
\end{equation}
Since this is actually a coupled system of equations, it is easier to work in terms of the total magnitude of $\Delta \bm{u}$,
\begin{equation}
    U \equiv \lVert \Delta \bm{u} \rVert = \left[\left(\Delta \bm{u}\right) \cdot \left(\Delta \bm{u}\right)\right]^{1/2}
\end{equation}
and
\begin{equation}
    \partial_t U = \frac{\left(\Delta \bm{u}\right) \cdot \partial_t \left(\Delta \bm{u}\right)}{\lVert \Delta \bm{u} \rVert}.
\end{equation}
Therefore, we have the intermediate equation
\begin{equation}
    \partial_t U = -\frac{6 \pi a \nu \rho}{m} \left(\frac{2 a}{\nu}\right)^{2/5} U^{7/5}.
\end{equation}
Using the solution to the intermediate equation in the original system, we find
\begin{equation}
    \Delta \bm{u}\!\left(t\right) = \left[\frac{5 m}{5 m + 12 \pi \rho t \left(4 a^7 \nu^3 \lVert \Delta \bm{u}_0 \rVert^2\right)^{1/5}}\right]^{5/2} \Delta \bm{u}_0.
    \label{eqn:stokesmed}
\end{equation}

\subsubsection{Large Reynolds number}

\noindent In the last subregime of Stokes drag, we have $\dragcoeff = \frac{44}{100}$, and the equations for the velocity differences are
\begin{equation}
    \partial_t \left(\Delta \bm{u}\right) = -\frac{11 \pi a^2 \rho}{50 m} \lVert \Delta \bm{u} \rVert \Delta \bm{u}.
\end{equation}
Using the same strategy as for intermediate Reynolds numbers, we have
\begin{equation}
    \partial_t U = -\frac{11 \pi a^2 \rho}{50 m} U^2.
\end{equation}
The final solution is given by
\begin{equation}
    \Delta \bm{u}\!\left(t\right) = \left[\frac{50 m}{50 m + 11 \pi a^2 \rho t \lVert \Delta \bm{u}_0 \rVert}\right] \Delta \bm{u}_0.
    \label{eqn:stokeslarge}
\end{equation}

\subsubsection{Transition points}

\noindent Since the solution is inherently piecewise, because of the piecewise definition of $\dragcoeff$, analytic expressions for the time at which transitions between Reynolds number subregimes are needed. For a given $\Delta \bm{u}_0$, a particle moves from $\Reynolds > 800$ to $1 < \Reynolds < 800$ at time
\begin{equation}
    t_{800} = \frac{m \left(a \lVert \Delta \bm{u}_0 \rVert - 400 \nu\right)}{88 \pi a^2 \nu \rho \lVert \Delta \bm{u}_0 \rVert},
    \label{eqn:t800}
\end{equation}
the time at which $\Reynolds = 800$ exactly. Similarly, a particle moves from $\Reynolds > 1$ to $\Reynolds < 1$ at time
\begin{equation}
    t_1 = \frac{5 m}{12 \pi a \nu \rho} \left[1 - \left(\frac{\nu}{2 a \lVert \Delta \bm{u}_0 \rVert}\right)^{2/5}\right].
    \label{eqn:t1}
\end{equation}
Equations~\ref{eqn:stokessmall}, \ref{eqn:stokesmed}, \ref{eqn:stokeslarge}, \ref{eqn:t800}, and \ref{eqn:t1} specify the evolution of velocity differences in the Stokes drag regime.

\subsection{Epstein regime}

\noindent The Epstein regime (except at high Mach numbers) is functionally similar to the small Reynolds number subregime of Stokes drag. The exponential solution is given by
\begin{equation}
    \Delta \bm{u}\!\left(t\right) = \exp\!\left[-\frac{4\pi a^2 \rho t}{3 m} \left(\frac{8 c_s^2}{\pi \gamma}\right)^{1/2}\right] \Delta \bm{u}_0.
\end{equation}

\section{Supplemental figures}
\label{adx:suppfig}
\restartappendixnumbering

\begin{figure*}
    \centering
    \includegraphics[width=\linewidth]{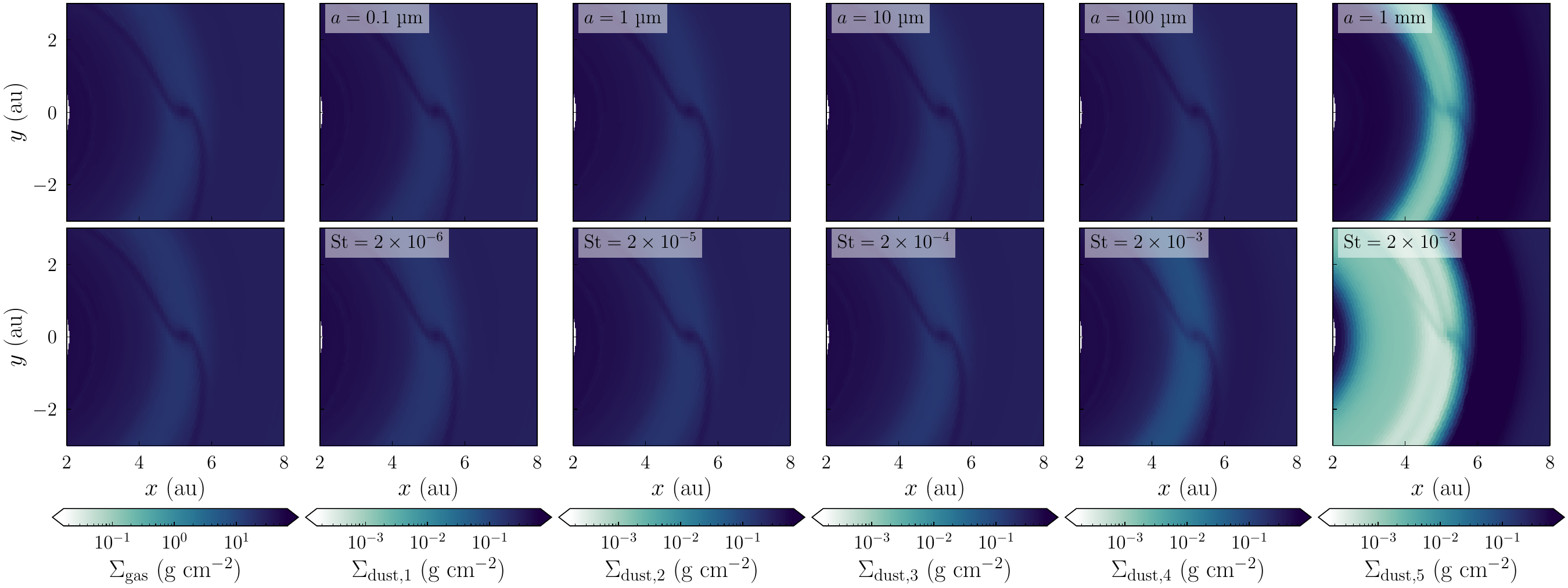}
    \caption{Same as Figure~\ref{fig:morphology-30earth}, but zoomed in around the planet and its gap.}
    \label{fig:morphology-30earth-zoom}
\end{figure*}

\begin{figure*}
    \centering
    \includegraphics[width=\linewidth]{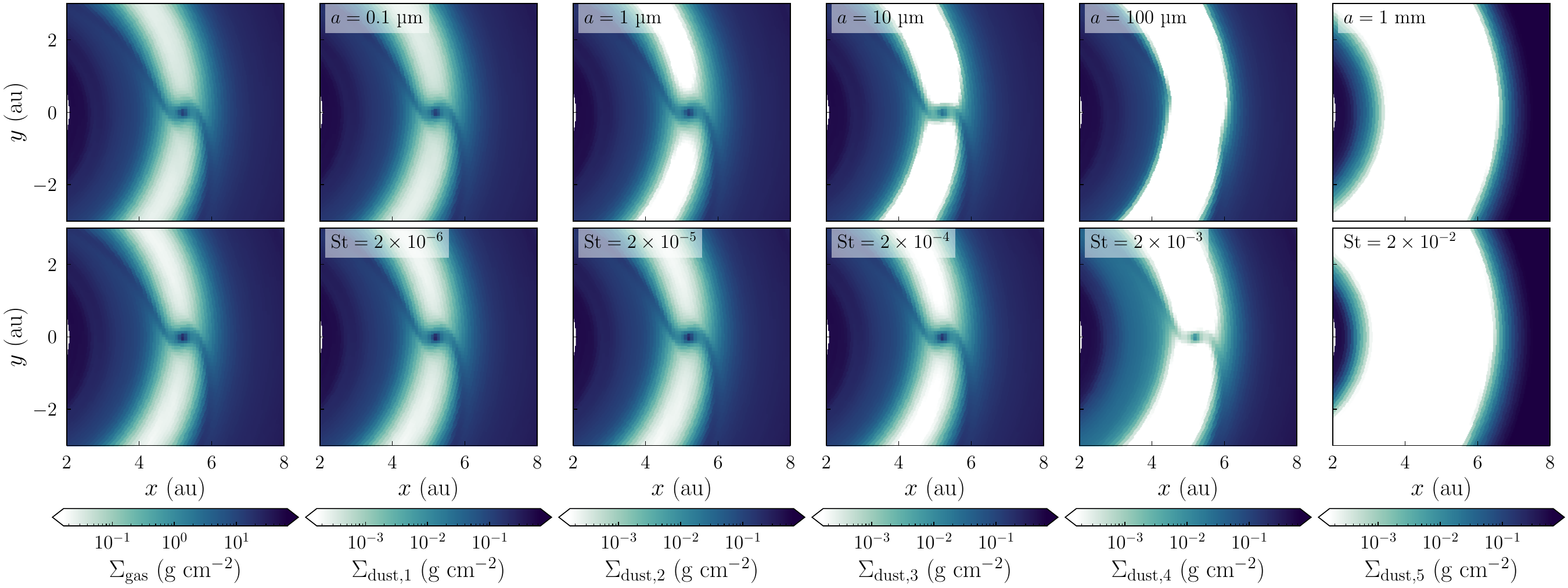}
    \caption{Same as Figure~\ref{fig:morphology-jupiter}, but zoomed in around the planet and its gap.}
    \label{fig:morphology-jupiter-zoom}
\end{figure*}

\noindent In Figures~\ref{fig:morphology-30earth-zoom} and \ref{fig:morphology-jupiter-zoom}, we show magnified versions of Figures~\ref{fig:morphology-30earth} and \ref{fig:morphology-jupiter}, respectively, to better show detail around the location of the planet and its gap.

\end{document}